# Calculating the submillimeter emissivity of dust grains in spiral galaxies

E. M. Xilouris

*National Observatory of Athens, I. Metaxa & Vas. Pavlou str., Palaia Penteli, 15236, Athens, Greece*

**Abstract.** We use the radiation transfer simulation of Xilouris et al. [1] to constrain the quantity of dust in three nearby spiral galaxies (NGC 4013, NGC 5907 and NGC 891). The predicted visual optical depth from the model is compared with the thermal continuum radiation detected from these galaxies at 850 $\mu$m. This comparison yields the emissivity of dust grains in the submillimeter waveband which is a factor 4 higher than the benchmark, semi-empirical model of Draine & Lee [2]. Our estimates are more closely aligned with recent measurements carried out in the laboratory on *amorphous* carbon and silicate particulates. A comparison between the distribution of 850 $\mu$m surface brightness and the intensity levels in the $^{12}$CO(1-0) and 21 cm lines underlines the spatial association between dust detected in the submillimeter waveband and molecular gas clouds. We suggest that the relatively high emissivity values that we derive may be attributable to amorphous, fluffy grains situated in denser gas environments.

## INTRODUCTION

The grain parameter allowing the FIR optical depth to be converted into dust column density and ultimately dust mass – the emissivity $Q$ – still remains uncertain by nearly an order of magnitude [3]. $Q$ indicates the efficiency with which dust grains of a particular temperature emit FIR thermal radiation. Due to a dearth of direct measurements, a large proportion of astronomers rely on a single estimate of emissivity at 125 $\mu$m obtained 20 years ago for a single Galactic reflection nebula [4]. Another work frequently cited is that of Draine & Lee [2]. These authors propose emissivities on the basis of laboratory experiments for $\lambda \leq 60$ $\mu$m and use primarily a model based on solid state theory to extrapolate to longer wavelengths. The 100 $\mu$m-brightness per H-atom, predicted by Draine & Lee [2] for high-lattitude dust clouds, agrees within 30% of the measurements carried out by the *Infrared Astronomical Satellite* (IRAS). However, as the authors themselves acknowledge (p.107 of [2]), the emissivity proposed for $\lambda > 300$ $\mu$m may be too low by a factor of 3-4 to be consistent with astronomical observations.

In this paper, we carry out a direct measurement of the submillimeter (submm) emissivity of dust grains situated in three nearby, quiescent spiral disks – NGC 4013, NGC 5907 and NGC 891. The technique, explained in detail in the next section, compares two tracers of what is believed to be the same dust grain population in order to infer the emissivity. The tracers in question are submm surface brightness and visual optical depth. Apart from estimating the submm emissivity we explore how the dust traced in thermal continuum emission relates to the various gas phases ($H_2$, HI) within the disk.

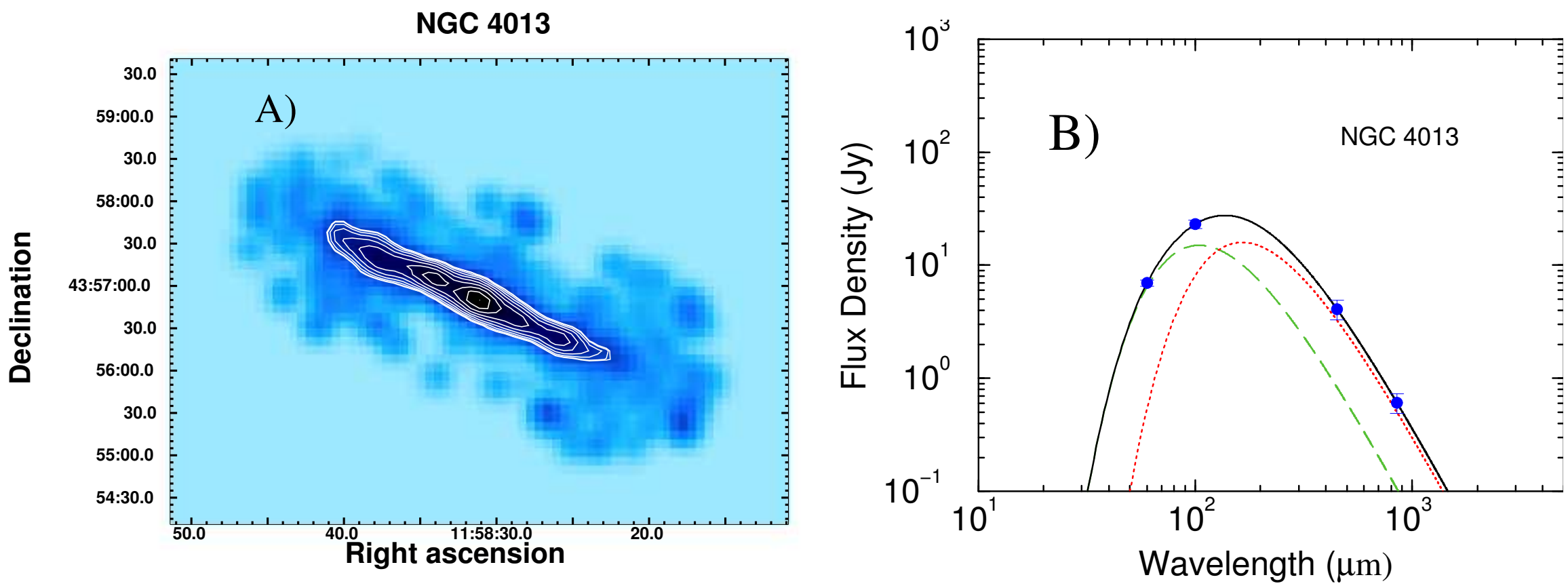

**FIGURE 1.** **A.** NGC 4013 observed in the 850 $\mu m$ continuum with SCUBA. Both the background image and the contours refer to emission at 850 $\mu m$ The contour levels are at 10.0, 10.4, 11.4, 13.3 15.9, 19.2, 23.3, 28.1, 33.7 and 40.0 mJy/16″ beam. **B.** Flux densities for NGC 4013 observed by IRAS (60 and 100 $\mu m$) and SCUBA (450 and 850 $\mu m$). The thermal spectrum has been fitted by the sum of a warm and cold dust component (dashed and dotted lines respectively). The total SED is given by the solid curve.

# TECHNIQUE FOR DETERMINING THE EMISSIVITY

Alton et al. [7] have already discussed the method in some detail but we summarise here the basic precepts. Classical dust grains of radius 0.1 $\mu$m will tend to reach an equilibrium temperature $T$ when immersed in a stellar radiation field. Under such circumstances, heating due to the absorption of optical, ultraviolet and near-infrared photons is exactly balanced by cooling due to emission of mid and far-infrared radiation. The efficiency with which grains emit submm radiation depends on their composition and structure and can be expressed by the emissivity $Q$, a dimensionless quantity which indicates how the flux density ($F$) recorded at wavelength $\lambda$ compares with that emitted by a blackbody:

$$F(\lambda) = \frac{n\sigma}{D^2} Q(\lambda) B(\lambda, T) \tag{1}$$

Here $D$ is the distance to a dust cloud containing $n$ dust grains of geometrical cross-section $\sigma$. $B$ is the planck function for a blackbody of temperature $T$.

In the past, a major obstacle has been constraining $B(\lambda, T)$ which, near the peak of the FIR spectrum ($\lambda \sim 200$ $\mu$m for spiral galaxies), is highly sensitive to the adopted grain temperature $T$. The temperatures we shall use, however, are derived from fitting the Rayleigh-Jeans tail of spectrum and in this regime $B \propto T$ approximately. Inspecting Eq. (1), we recognise that, even if $F$ and $B$ are known with relative certainty we are required to disentangle $n\sigma$ from $Q$. We achieve this by substituting the visual optical depth defined as follows:

$$\tau_V = N\sigma Q(V) \tag{2}$$

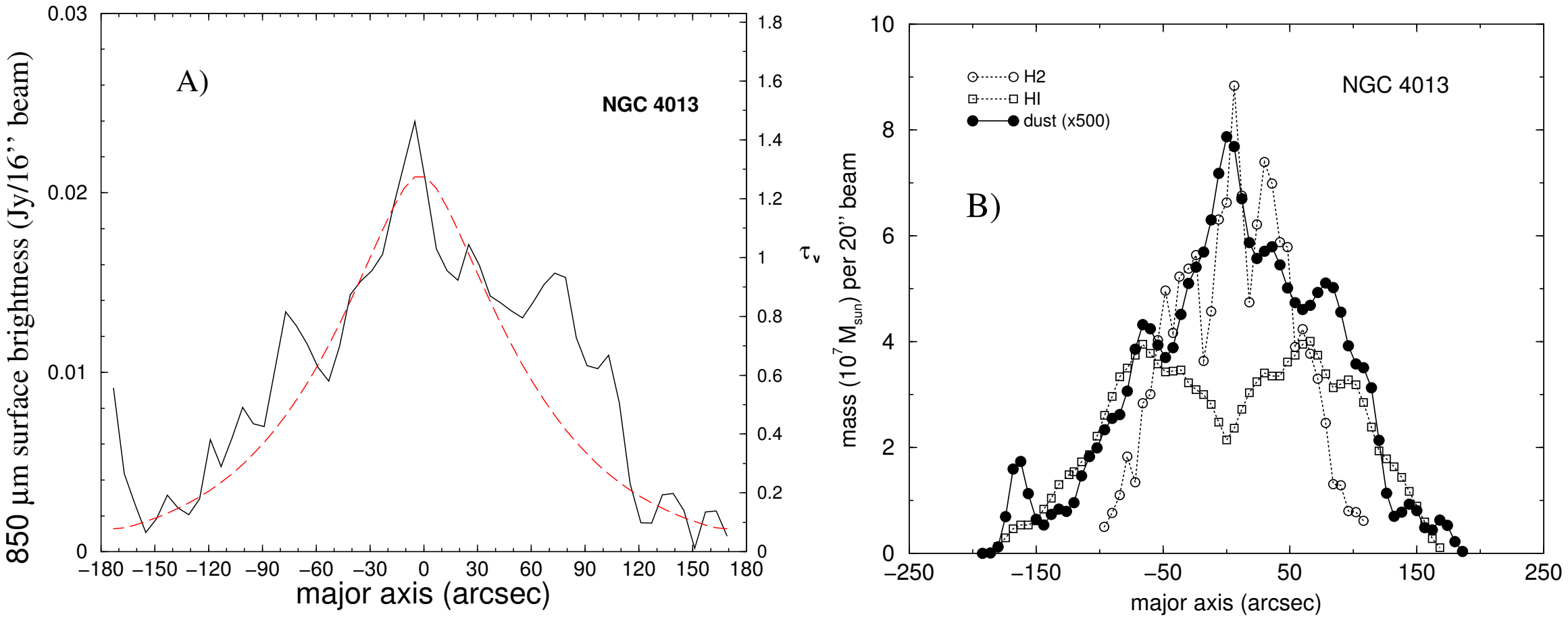

**FIGURE 2.** **A.** Major axis profiles of NGC 4013 in 850 $\mu m$ surface brightness (solid line, left axis) and visual optical depth (dashed line, right axis). The latter has been inferred from radiation transfer modelling and the map in visual optical depth has been smoothed to the corresponding spatial resolution of the submm data (16″) before profiling. **B.** Dust and neutral gas profiles along the major axis of NGC 4013. The solid circles denote the distribution of grains (using a multiplicative factor of 500 to clarify the plot). The open circles and open squares trace the respective distributions of molecular and atomic gas within the disk.

where $Q(V)$ is the extinction efficiency in the V-band ($\lambda = 0.55\,\mu$m) and $N$ is the number of grains per unit area on the sky. At the same time we substitute the surface brightness $f(\lambda)$ for the flux density $F(\lambda)$ so that the distance to the dust cloud $D$ is eliminated. This re-formulation of Eq. (1) yields:

$$\frac{\tau_V}{f(\lambda)} = \frac{Q(V)}{Q(\lambda)} \frac{2.2 \times 10^{-18}}{B(\lambda, T)} \tag{3}$$

In general, it is the ratio $\frac{Q(V)}{Q(\lambda)}$ which is of immediate interest in comparing our results with other studies. However, if we are to quantify the number of FIR-emitting grains in Eq. (1), and thereby ultimately determine the dust mass of the system, we should derive the absolute emissivity $Q(\lambda)$. To do this we must specify $Q(V)$, $B(\lambda,T)$, $f(\lambda)$ and $\tau_V$ in Eq. (3). The first of these quantities, $Q(V)$, is believed to lie, with some certainty between 1 and 2 ([8], (p. 60); [7]) and we assume a value of 1.5 hereafter. The submm surface brightness, $f(\lambda)$, is a quantity that comes directly from the observations (Figure 1A) while the temperature associated with the blackbody $B(\lambda,T)$ follows from the shape of the FIR spectrum (Figure 1B). The optical depth, $\tau_V$, is derived by applying a realistic 3D radiative transfer model to the optical observations of these galaxies [1].

After smoothing the map in V-band optical depth to the required spatial resolution we profiled both tracers ($f(\lambda)$ and $\tau_V$) along the major axis of the galaxy in order to derive the mean ratio $\frac{\tau_V}{f(\lambda)}$ (required for Eq. (3)). These profiles are illustrated in Figure 2A for the case of NGC 4013.

## DISCUSSION ON THE EMISSIVITY (Q)

Using our technique which was described in the previous section we infer emissivity values at $850\mu m$ of $1.07 \times 10^{-4}$, $1.25 \times 10^{-4}$ and $9.37 \times 10^{-5}$ for the galaxies NGC 4013, NGC 5907 and NGC 891 respectively. These values are a factor of $\sim 4$ higher than the values of Draine & Lee [2] which are widely used. Our results lie within the range suggested by recent tests on *amorphous* carbon and silicate particulates conducted in the laboratory [5], [6].

Our central assumption is that grains extinguishing optical/NIR radiation in spiral disks will also be the main emitters of submm thermal radiation. Hildebrand [4] has already shown that the dependencies of optical extinction and FIR emission on grain size are broadly similar. Using the MRN grain-size distribution as a test-case [9] it is shown that, for the same grain-size distribution in all parts of the ISM, grains of roughly the same size will dominate both optical extinction and submm thermal emission (see Appendix of [10]). The enhanced emissivity we measure in the previous section, however, may suggest that our SCUBA images are sensitive to a population of grains that are significantly larger than the classical $0.1\mu$m grains dominating optical extinction. This question of bi-modality in the grain size can only be solved expeditiously by a decomposition of galactic submm emission into discrete and diffuse sources (as has been carried out for the Milky Way, at shorter wavelengths, using IRAS). Prior subtraction of discrete sources from our SCUBA images, in order to eliminate larger, non-classical grains, would act to *lower* the emissivity we derive for the classical population dominating the optical extinction.

The submm emission detected from our objects follows closely the distribution of molecular gas, the latter being traced in the $^{12}$CO(1-0) line (Figure 2B). A physical explanation for the relatively high emissivity values we derive might be that dust emitting strongly in the submm waveband is situated chiefly in molecular gas clouds where the elevated density is conducive to the formation of amorphous, fluffy grains. Such grains are expected to possess emissivity values that are a factor 3 or so higher than refactory cores circulating in the diffuse ISM [11].